\documentclass[12pt]{article}
\usepackage{mydef}
\usepackage{amssymb}
\usepackage[dvips]{graphicx}
\usepackage{feynmf}
 \usepackage{mathrsfs}

\usepackage{epsfig}
\usepackage{array}
\usepackage{float}

\usepackage{a4}
\usepackage{epsfig}
\usepackage{a4wide}

\def\ra{\rightarrow} 
\newcommand{\beq}{\begin{equation}}
\newcommand{\eeq}{\end{equation}}
\def\gs{\mathrel{ \rlap{\raise
0.511ex \hbox{$>$}}{\lower 0.511ex \hbox{$\sim$}}}} \def\ls{\mathrel{
\rlap{\raise 0.511ex \hbox{$<$}}{\lower 0.511ex \hbox{$\sim$}}}}

\newcommand{\dms}{\mbox{$\Delta m^2_{\odot}$ }}
\newcommand{\dma}{\mbox{$\Delta m^2_{\rm A}$ }}

 \def\ra{\rightarrow}

\def\gtap{\mathrel{ \rlap{\raise 0.511ex \hbox{$>$}}{\lower 0.511ex
   \hbox{$\sim$}}}} \def\ltap{\mathrel{ \rlap{\raise 0.511ex
   \hbox{$<$}}{\lower 0.511ex \hbox{$\sim$}}}}

   \newcommand{\betabeta}{\mbox{$(\beta \beta)_{0 \nu} $}}

   \newcommand{\meff}{\mbox{$\left| < \! m \! > \right|$}}
   
   \newcommand{\hbeta}{$\mbox{}^3 {\rm H}$ $\beta$-decay }

\newcommand{\am}{\alpha}
\newcommand{\an}{\beta}

\begin{document}
\hfill{Ref. SISSA 1/2003/EP} 
\rightline{UCLA/03/TEP/2}
\rightline{January 2003}
\rightline{hep-ph/0301xxx}

\begin{center}
{\bf Quasi-Degenerate Neutrino Mass Spectrum, 
$\mu \ra e +\gamma$ Decay  and Leptogenesis\\
}

\vspace{0.4cm}
S. Pascoli~$^{a)}$, \hskip 0.2cm S. T. Petcov~$^{b,c)}$
\footnote{Also at: Institute of Nuclear Research and Nuclear Energy,
Bulgarian Academy of Sciences, 1784 Sofia, Bulgaria} ~and~
C. E. Yaguna~$^{b)}$

\vspace{0.2cm}

{\em $^{a)}$Department of Physics, University of California, 
Los Angeles CA 90095-1547, USA\\ }
\vspace{0.2cm} {\em $^{b)}$Scuola Internazionale Superiore di Studi
Avanzati, I-34014 Trieste, Italy\\ }
\vspace{0.2cm} {\em $^{c)}$Istituto Nazionale di Fisica Nucleare,
Sezione di Trieste, I-34014 Trieste, Italy\\ }

\end{center}

\begin{abstract}
In a large class of SUSY GUT models
with see-saw mechanism of 
neutrino mass generation, lepton 
flavor violating (LFV) decays 
$\mu \rightarrow e + \gamma$,
$\tau \rightarrow \mu + \gamma$, etc.,
are predicted with 
rates that are within the reach 
of present and planned experiments. 
A crucial element in these 
predictions is the matrix of neutrino Yukawa 
couplings $\ynu$ which can be expressed
in terms of the light and RH heavy 
neutrino masses, the
neutrino mixing PMNS matrix $U$,  
and an orthogonal 
matrix $\mathbf{R}$.
Leptogenesis can take place only if 
$\mathbf{R}$ is complex.
Considering the case of 
quasi-degenerate neutrinos
and assuming that $\mathbf{R}$ 
is complex, we derive 
simple analytical 
expressions for the 
$\mu \rightarrow e + \gamma$,
$\tau \rightarrow \mu + \gamma$ and 
$\tau \rightarrow e + \gamma$
decay rates. Taking into account the 
leptogenesis constraints
on the relevant parameters
we show that the predicted rates of the
LFV decays $\mu \rightarrow e + \gamma$,
and $\tau \rightarrow e + \gamma$
are generically enhanced by  
a factor of
$\sim 10^{3}$ to $\sim 10^{6}$ 
with respect to the rates calculated
for real $\mathbf{R}$, while
the $\tau \rightarrow \mu + \gamma$
decay rate is enhanced approximately by
two orders of magnitude. 

\end{abstract}

\newpage

\section{Introduction}
\vspace{-0.2cm}

\hskip 1truecm  
   The solar neutrino experiments
\cite{Cl98,SKsol,SNO1,SNO2}, 
the data on atmospheric neutrinos
obtained by the Super-Kamiokande 
collaboration \cite{SKatm00}, and 
the results from the KamLAND 
reactor antineutrino 
experiment \cite{KamLAND},
provide very strong evidences for
mixing and oscillations
\cite{Pont57,Pont67,GPont69} of flavour neutrinos. 
The evidences for solar 
$\nu_e$ oscillations into active neutrinos $\nu_{\mu,\tau}$,
in particular, were spectacularly reinforced 
by the combined Super-Kamiokande and 
first SNO \cite{SNO1} data, 
by the more recent SNO data \cite{SNO2},
and by the just published 
first results of the KamLAND 
\cite{KamLAND} experiment. 

  The interpretation of the solar and
atmospheric neutrino, and of the KamLAND
data in terms of 
neutrino oscillations requires
the existence of 3-neutrino mixing
in the weak charged lepton current 
(see, e.g., \cite{BGG99,SPWIN99}):
\begin{equation}
\nu_{l \mathrm{L}}  = \sum_{j=1}^{3} U_{l j} \, \nu_{j \mathrm{L}}~.
\label{3numix}
\end{equation}
\noindent Here $\nu_{lL}$, $l  = e,\mu,\tau$,
are the three left-handed flavor 
neutrino fields,
$\nu_{j \mathrm{L}}$ is the 
left-handed field of the 
neutrino $\nu_j$ having a mass $m_j$
and $U$ is the Pontecorvo-Maki-Nakagawa-Sakata (PMNS)
neutrino mixing matrix \cite{Pont57}. 
It follows from the results of the
\hbeta experiments \cite{MoscowMH3} that 
$m_{j} < 2.2$ eV.
The existence of the flavour 
neutrino mixing, eq. (\ref{3numix}),
implies that the individual lepton charges,
$L_e,~L_{\mu}$ and $L_{\tau}$ are not conserved
(see, e.g., \cite{BiPet87}).
Therefore, lepton flavour violating (LFV)  
processes like $\mu \rightarrow e + \gamma$,
$\mu^{-} \rightarrow e^{-} + e^{+} + e^{-}$, 
$\tau \rightarrow \mu + \gamma$, 
$\mu^{-} + (A,Z) \rightarrow e^{-} + (A,Z)$,
are allowed. However, if 
the neutrino (lepton) mixing
in the weak charged lepton current is the only
source of $L_e,~L_{\mu}$ and $L_{\tau}$ 
non-conservation, 
as in the minimally extended Standard
Theory with massive neutrinos,
the rates and cross-sections of
the LFV processes
are suppressed by the factor \cite{SP76}
$(m_{j}/M_W)^4 < 5.6\times 10^{-43}$,
$M_W$ being the $W^{\pm}$ mass,
which renders them unobservable.

  The experimentally suggested smallness of the 
neutrino masses can naturally 
be explained by the see-saw mechanism 
of neutrino mass generation \cite{saw}.
The see-saw mechanism requires 
the existence of heavy right-handed (RH)
Majorana neutrinos. Right-handed neutrinos 
\cite{Pont67,BiPont76} 
are completely neutral under 
the Standard Theory gauge symmetry group. 
Consequently, they can acquire 
Majorana masses $M_R$ that are not related 
to the electroweak symmetry breaking mechanism, 
and can, in principle, be much heavier 
than any of the known particles. 
The heavy RH Majorana neutrinos  
can generate
through their CP-violating decays
the observed baryon 
asymmetry of the Universe \cite{FukYan}.
In grand unified theories (GUT)
their masses are typically  
by a few to several orders of magnitude 
smaller than the scale of unification
of the electroweak and strong interactions,
$M_{X} \sim 2\times 10^{16}$ GeV.
However, their presence  
in a theory can lead to a severe 
hierarchy problem 
associated with the existence of two very 
different mass (energy) scales: the electroweak 
symmetry breaking and 
the RH Majorana mass scale. 
In supersymmetric (SUSY) GUT theories 
the hierarchy between these two mass scales
is stabilized.
Hence, the SUSY GUT theories 
incorporating the see-saw mechanism 
of neutrino mass generation
provide a consistent 
and appealing 
framework to account for
neutrino masses and for the 
baryon asymmetry in the Universe. 

   SUSY theories have additional 
sources of lepton charge non-conservation. 
In spite of the possible flavor-blindness 
of SUSY breaking, the supersymmetrization 
of the see-saw mechanism,
for instance, can induce new LFV
effects \cite{mas}. If SUSY is 
broken above the RH Majorana mass scale, 
as, e.g, in gravity-mediated breaking scenarios, 
there are renormalization group effects that 
generate new 
lepton charge non-conserving
couplings at low energy even if
such couplings are absent 
at the GUT scale. 
In contrast to the non-supersymmetric case, 
these couplings give contributions 
to the amplitudes of the LFV 
decays and reactions  
which are not suppressed by 
the small values of  
neutrino masses
and the LVF processes can proceed with
rates and cross-sections which
are within the sensitivity of 
presently operating   
and proposed experiments 
(see, e.g., \cite{Kuno99}).

  The solar and atmospheric neutrino
data and the data from  
the reactor $\bar{\nu}_e$ experiments
KamLAND, CHOOZ and Palo Verde, 
were used successfully 
for determining the pattern 
of the $3-\nu$ mixing
and the values of 
the two independent neutrino 
mass-squared differences,
$\dms$ and $\dma$, which 
drive the solar and atmospheric 
neutrino oscillations.
Under the rather plausible 
assumption of CPT-invariance, 
for instance,
the recent KamLAND results practically 
establish \cite{KamLAND} the
large mixing angle (LMA)
MSW solution as unique solution
of the solar neutrino problem,
with $\dms \sim 7\times 10^{-5}~{\rm eV^2}$
and $\tan^2\theta_{\odot} \sim 0.40$
favored by the data,
$\theta_{\odot}$ being the mixing angle which
controls the solar $\nu_e$ oscillations.
The analyses of the atmospheric 
neutrino data show
that $\dma$ and the mixing parameter
$\sin^22\theta_{\rm A}$, 
responsible for the dominant atmospheric 
$\nu_{\mu}$ ($\bar{\nu}_{\mu}$) oscillations into
$\nu_{\tau}$ ($\bar{\nu}_{\tau}$), 
have values
$\dma \sim 3\times 10^{-3}~{\rm eV^2} \gg \dms$,
and $\sin^22\theta_{\rm A} \sim (0.9 - 1.0)$.
The existing data, however,
does not allow to determine 
the sign of $\dma$. Furthermore,
the neutrino oscillations
are not sensitive to the absolute values 
of neutrino masses.
Correspondingly, there are 
three different 
types of 3-neutrino mass spectra 
which are compatible with the 
existing neutrino oscillation data
\cite{SPAS94} (see also, e.g., \cite{BPP1}): 
normal hierarchical
(NH), $m_{1}\ll m_{2}\ll m_{3}$, 
inverted hierarchical (IH),
$m_{1}\ll m_{2}\cong m_{3}$, and
quasi-degenerate (QD),
$m_{1}\simeq m_{2}\simeq m_{3}$, $m^2_{1,2,3} \gg \dma$.

   In the case of QD spectrum, 
neutrino masses can be measured directly in the 
\hbeta experiments
which are sensitive to the 
$\bar{\nu}_e$ mass, $m_{\bar{\nu}_ e} \cong m_{1,2,3}$.
The present bound
obtained in these experiments 
reads \cite{MoscowMH3}, 
$m_{1,2,3} \cong m_{\bar{\nu}_ e} < 2.2$ eV.
Sensitivity  
to values of $m_{1,2,3} \simeq 0.35$ eV 
are planned to be reached in the KATRIN
experiment \cite{KATRIN}. 
If the massive neutrinos $\nu_j$ 
are Majorana particles, as is predicted by the
see-saw mechanism,
neutrinoless double-beta decay experiments 
can also provide information on the type of 
the neutrino mass spectrum and 
on the absolute neutrino mass scale
(see, e.g., \cite{BGGKP99,BPP1} and 
the references quoted therein).
They measure a combination of masses 
and mixing parameters 
known as the effective Majorana mass parameter, 
$\meff$ (see, e.g., \cite{BiPet87,BPP1}). 
The most stringent constraints  
on $\meff$ were obtained in the $^{76}$Ge experiments:
$\meff <0.35$ eV \cite{76Ge00} ($90\%$ C.L.), and 
$\meff < (0.33 - 1.35)$ eV \cite{IGEX00} ($90\%$ C.L.).
Higher, or considerably higher, sensitivities
to $\meff$ are planned to be achieved
in several \betabeta-decay experiments 
of the next generation (for a review see, e.g., \cite{MSpironu02}).
If neutrinos have a QD mass spectrum, 
they can be relevant cosmologically
through their contribution to the hot 
dark matter component of the Universe.
The sum of neutrino masses 
$(m_1 + m_2 + m_3)$
can be determined with a precision 
of $\sim (0.04 - 0.10)$ eV
from cosmological and astrophysical 
data \cite{Hu99}.

  The Universe seems to be made only of matter; 
cosmologically significant amounts of 
antimatter have never been observed.
This asymmetry between matter and antimatter 
can be understood as the result of the 
dynamical evolution of an initially symmetric 
Universe in which baryon number is not conserved, 
\emph{C}- and \emph{CP}- symmetries
are violated and a deviation from 
thermal equilibrium exists \cite{Sakh}. 
If these conditions are fulfilled, 
baryogenesis, the process which generates 
an excess of baryons over antibaryons, 
can take place. At present, one of 
the most favored 
scenarios for baryogenesis is 
the leptogenesis scenario \cite{FukYan} 
in which the 
heavy RH neutrinos play a fundamental role. 
Their \emph{CP}-violating and out-of-equilibrium 
decays produce a lepton asymmetry that is 
partially converted into a baryon asymmetry 
through anomalous electroweak processes. 
Leptogenesis has the attractive feature 
of providing a link between neutrino masses 
and the baryon asymmetry.

   In a large class of SUSY GUT models
with see-saw mechanism of 
neutrino mass generation
and flavour-universal soft SUSY 
breaking at the GUT scale (see, e.g., \cite{iba,JohnE}), 
the LFV processes 
and leptogenesis are 
related: they both depend 
(although in different ways)
on the matrix of neutrino 
Yukawa couplings $\ynu$.
The latter is 
one of the basic ingredients of 
the see-saw mechanism.
The matrix $\ynu$ can be expressed
in terms of the light neutrino and heavy RH 
neutrino masses, the
neutrino mixing PMNS matrix $U$,  
and an orthogonal matrix $\mathbf{R}$.
Leptogenesis can take place only if 
$\mathbf{R}$ is complex.
Working in the framework
of the indicated class of theories 
and taking $\mathbf{R}$ to be complex,
we derive in the present 
article simple analytical expressions 
for the $\mu \rightarrow e + \gamma$,
$\tau \rightarrow \mu + \gamma$ and 
$\tau \rightarrow e + \gamma$
decay rates in the case of 
quasi-degenerate neutrino mass spectrum.
We use the model of leptogenesis of \cite{kume}, 
in which the heavy RH neutrinos are produced
non-thermally in inflaton decays, 
to obtain constraints on the 
parameters which determine the 
leading contribution in 
the LFV decay rates. 
Taking into account the 
leptogenesis constraints, 
we show that the rates of the
LFV decays $\mu \rightarrow e + \gamma$,
$\tau \rightarrow \mu + \gamma$ and
$\tau \rightarrow e + \gamma$,
obtained for complex $\mathbf{R}$,
are generically strongly  enhanced 
with respect to those calculated
for real $\mathbf{R}$. 
We present quantitative results for 
the enhancement factors 
for the indicated three LFV decays.

  Detailed predictions for 
the rates of the LFV processes in
the class of SUSY GUT models
with see-saw mechanism
considered in our work
were obtained, e.g., in 
refs. \cite{iba,JohnE,tan,Saclay,DepPas}.
The case of quasi-degenerate neutrinos 
we analyze was discussed, in particular,
in \cite{iba,tan,DepPas}. However,
the results in these articles were obtained
for real matrix $\mathbf{R}$.
In \cite{Saclay} the case of 
hierarchical neutrino mass spectrum 
was considered.
The articles quoted in ref. \cite{JohnE}
contain rather comprehensive
study of the LFV processes, 
including the case of complex
$\mathbf{R}$ and the leptogenesis
constraints, but for
light neutrino mass spectra 
with normal and with inverted hierarchy.

\vspace{-0.4cm}
\section{The neutrino Yukawa coupling}\label{uno}
\vspace{-0.2cm}
 
\hskip 1truecm The superpotential of the lepton sector 
in the MSSM with RH neutrinos is given by:
\beq
W_{lepton}=\hat{l}_{\sss L}^{c~T}~\ye  
\hat{L}\cdot \hat{H}_d + \hat{N_L}^{c~T}\ynu 
\hat{L}\cdot \hat{H}_u -
\frac 12\hat{N}_L^{c\,T}\mathbf{M_{\sss R}}\hat{N}_L^c \;,
\label{W}
\eeq
%
where the family indices were suppressed.
Here $\hat{L}_j$, $j=e,\mu,\tau \equiv 1,2,3$, 
represent the chiral 
super-multiplets of the $SU(2)_L$ doublet 
lepton fields, 
$\hat{l}_{\sss j L}^{c}$, $j=e,\mu,\tau \equiv 1,2,3$,
is the super-multiplet of the $SU(2)_L$ singlet 
lepton field $l_{\sss j L}^{c} \equiv C\bar{l}_{\sss j R}^{T}$,
where $C$ is the charge conjugation matrix and 
$l_{\sss j R}$ is the
right-handed charged lepton field,
$\hat{N}_{jL}^c$ is the super-multiplet 
of the $SU(2)_L$ singlet 
neutrino field  $N_{jL}^c \equiv C\bar{N}_{\sss j R}^{T}$,
where $N_{\sss j R}$ is the RH neutrino field, and
$\hat{H}_u$ and $\hat{H}_d$ are the
super-multiplets of the two Higgs doublet fields
$H_u$ and $H_d$ carrying weak hypercharges 
$-\frac 12$ and $\frac 12$, respectively. 
In eq. (\ref{W}), $\ynu$ is the $3\times 3$ matrix of 
neutrino Yukawa couplings, 
$\ye$ is the $3\times 3$ matrix of 
the Yukawa couplings of the charged 
leptons, and $\mathbf{M_{\sss R}}$ is the 
Majorana mass matrix of the RH neutrinos $N_{\sss j R}$.
We can always choose a basis in which both
$\ye$ and  $\mathbf{M_{\sss R}}$ are diagonal. 
We will work in that basis 
and will denote by $D_M$ the corresponding 
diagonal RH neutrino mass matrix,
$D_M = diag(M_1,M_2,M_3)$.

 The see-saw mechanism generates    
a Majorana mass matrix for the left-handed neutrinos of the form:
\beq
\mathbf{m}_\nu = (\ynu v_u)^T D_M^{-1}(\ynu v_u),
\eeq
%
where $v_u$ is the vacuum expectation value 
of $H_u$. The neutrino mass matrix $\mathbf{m}_\nu$ 
is diagonalized by a single unitary matrix $U$ 
according to
\beq
D_m=U^T\mathbf{m}_\nu U\equiv diag(m_1,m_2,m_3)\;,
\eeq
%
where $U$ is the PMNS matrix in
the weak charged lepton current, eq. (\ref{3numix}).
%

  It is convenient to choose $m_j > 0$, to number
the massive neutrinos in such a way that
$m_1 < m_2 < m_3$, and to work with 
Majorana neutrino fields $\nu_j$ 
which satisfy the Majorana condition:
$C(\bar{\nu}_{j})^{T} = \nu_{j},~j=1,2,3$. 
In this case the PMNS matrix U can be written as
\beq
U=V\cdot diag(1,e^{i\alpha},e^{i\beta}),
\label{UPMNS}
\eeq
%
where $\alpha$ and $\beta$
are two Majorana \emph{CP}-violating phases 
\cite{BHP80}.
For $V$ one can use the standard parametrization
\beq \label{VPMNS}
V=\left(\ba{ccc} c_{13}c_{12}&c_{13}s_{12}&s_{13}e^{-i\delta}\\
-c_{23}s_{12}-s_{23}s_{13}c_{12}e^{i\delta}& 
c_{23}c_{12}-s_{23}s_{13}s_{12}e^{i\delta}&s_{23}c_{13}\\ 
s_{23}s_{12}-c_{23}s_{13}c_{12}e^{i\delta}& 
-s_{23}c_{12}-c_{23}s_{13}s_{12}e^{i\delta}&c_{23}c_{13}\ea\right)\,,
\eeq
%
with the usual notations, $s_{ij} \equiv \sin \theta_{ij}$,
etc. If, for instance, $\dms = \Delta m^2_{21}$
(neutrino mass spectrum with normal hierarchy) and 
$\dma = \Delta m^2_{31}$, one can identify
$\theta_{12} = \theta_{\odot}$, $\theta_{23} = \theta_{\rm A}$, 
while $\theta_{13}$ is limited by the data from
the CHOOZ and Palo Verde experiments \cite{CHOOZ,fogli},
$\sin^2\theta_{13} < 0.05$. 

  The matrix $D_m$ can be expressed as
\beq
D_m = U^T\ynu^Tv_uD_M^{-1}\ynu v_uU = 
U^T\ynu^T v_uD_M^{\sss -1/2} D_M^{\sss -1/2}\ynu v_u U.
\label{DmR}
\eeq
%
Following ref. \cite{iba},
we define the complex matrix $\mathbf{R}$:
\beq
\mathbf{R}\equiv D_M^{\sss -1/2}\ynu v_uUD_m^{\sss -1/2}~.
\label{R}
\eeq
%
  Given $D_M^{\sss 1/2}$, $D_m^{\sss 1/2}$ and $U$,
the most general neutrino Yukawa coupling matrix 
reads
\beq
\ynu=\frac{1}{v_u}D_M^{\sss 1/2} \mathbf{R}D_m^{\sss 1/2}U^\dagger\;.
\label{yuk}
\eeq 
%
 
 It follows from eq. (\ref{DmR}) that
$\mathbf{R}$ is an orthogonal matrix, 
\mbox{$\mathbf{R}\mathbf{R}^T=\mathbf{1}$}. 
In order for the 
leptogenesis scenario of baryon 
asymmetry generation to work,
$\mathbf{R}$ must be complex
and we will
keep $\mathbf{R}$ complex 
throughout this study.
As we will see, apart 
from being a 
necessary condition 
for leptogenesis, 
this leads also to   
drastically different predictions for
the rates of the LFV processes like
$\mu \rightarrow e + \gamma$, 
 $\tau \rightarrow e + \gamma$,
$\tau \rightarrow \mu + \gamma$.

The see-saw model  contains 
18 physical parameters - 
6 phases and 12 moduli. These include,
in the basis we work,
the 3 (real) masses of the 
heavy RH Majorana neutrinos, and 
9 moduli and 6 phases of $\ynu$
(3 of the 9 phases in $\ynu$ can be eliminated 
through a rephasing of the LH charged 
lepton fields).
At low energies it is convenient to
parametrize the model 
by the 3 angles and 3 phases 
of the PMNS mixing matrix
$U$, eqs. (\ref{UPMNS}) and (\ref{VPMNS}),
the 3 light neutrino masses, $m_{1,2,3}$,
and the 6 parameters - 3 moduli and 3 phases,
of the complex orthogonal matrix 
$\mathbf{R}$. 
The 3 additional real parameters
of the model are 
the 3 heavy RH Majorana neutrino masses,
contained in $D_M$.

       The 3 lepton mixing angles 
in the PMNS matrix $U$
and the two neutrino 
mass squared differences,
$\dms$ and $\dma$,
can be measured
with a relatively high precision
in neutrino oscillation experiments.
These experiments could also provide 
information on the
Dirac CP-violating phase $\delta$,
whereas information on the two 
Majorana CP-violating phases,
$\alpha$ and $\beta$,
can be obtained, in principle,  
in processes in which the Majorana 
nature of neutrinos manifests itself, 
such as \betabeta-decay, 
$K^{-}\rightarrow \pi^{+} + \mu^{-} + \mu^{-}$ decay, 
etc. (see, e.g., \cite{PPW,ABRabi02}).
The measurement of the neutrino mixing 
parameters would be complete with 
the determination of
the type of the neutrino mass spectrum 
and of the absolute neutrino mass scale.

 The probabilities of the LFV processes 
and the baryon asymmetry in leptogenesis 
depend on the see-saw parameters 
respectively via the quantities
\beq
\ynu^\dagger\ynu =      
\frac{1}{v^2_u} 
UD_m^{\sss 1/2}\mathbf{R}^\dagger\
D_M \mathbf{R}D_m^{\sss 1/2}U^\dagger\ ,
\eeq
%
and 
\beq
\mathrm{Im}\left[ (\ynu \ynu^\dagger)_{ij} \right]^2 =
\frac{1}{v^2_u}
\mathrm{Im}\left[ (D_M^{\sss 1/2}\mathbf{R}~D_m~\mathbf{R}^\dagger\ 
D_M^{\sss 1/2})_{ij} \right ]^2,~~i\neq j. 
\eeq
%
Thus, the matrix $\mathbf{R}$ enters into both 
the expressions for the rates of the LFV processes
and for the baryon asymmetry. 
 
We will consider in what follows the case of 
quasi-degenerate neutrino mass spectrum,
$m_1\cong m_2\cong m_3$, $m^2_{1,2,3} >> \dma,\dms$.
We can then write 
\be
m_1\equiv m_\nu\,,
\quad m_2 = 
m_\nu+\frac{1}{2m_\nu}\Delta m_\odot^2\,,
\quad m_3 = 
m_\nu+\frac{1}{2m_\nu}\dma\,,
\label{QDS}
\ee
%
where $m_\nu$ is the neutrino mass
determining the absolute neutrino mass scale
which is not known,
$m_\nu < 2.2$ eV \cite{MoscowMH3}.
It is natural to assume that 
also $D_M$ has quasi-degenerate eigenvalues,
$M_{1,2,3} \cong M_R$, 
$D_M \cong M_R\mathbf{1}$; otherwise, an 
exceptional fine-tuning between $\ynu$ and $D_M$ 
would be needed in order to obtain a 
QD spectrum for the light neutrinos. 

   In the case of QD neutrino mass spectrum one has
\be
\ynu \cong \frac{1}{v_u}M_R^{\sss 1/2}m_\nu^{\sss 1/2}
\mathbf{R} diag(1, 1 + \dms/(4m_\nu^2), 1 + \dma/(4m_\nu^2)) 
U^\dagger \cong
\frac{1}{v_u}M_R^{\sss 1/2}m_\nu^{\sss 1/2} 
\mathbf{R}U^\dagger \,.  
\label{YnuQD}
\ee
%
Hereafter corrections  $\mathcal{O}(\dms/(2m_\nu^2))$
and $\mathcal{O}(\dma/(2m_\nu^2))$
will be neglected.

   The matrix $\R$ can be parametrized as
\be
\R=e^{i\mathbf{A}}\mathbf{O}\,,
\ee
%
where $\mathbf{A}$ and $\mathbf{O}$ 
are real matrices. 
The orthogonality 
of $\R$ implies that $\mathbf{O}$ 
is orthogonal and 
$\mathbf{A}$ is antisymmetric.
A different parametrization of $\R$ 
has been used in previous 
works (see, e.g., \cite{iba}), 
but this one is particularly useful 
if the neutrino mass spectrum
is of the QD type. 
Up to corrections of the order of 
$\dma/(2m_\nu^2)$ and
$\dms/(2m_\nu^2)$, i.e.,
in the approximation of 
exact degeneracy of the three 
Majorana neutrinos $\nu_{1,2,3}$, 
the matrix $\mathbf{O}$
can be absorbed in 
the PMNS matrix $U$ - the latter
is defined up to a
real orthogonal matrix and
$U$ and $U\mathbf{O}$ lead to 
the same physics \cite{BrancoQD}.  
Thus, up to relatively 
small corrections,
$\mathbf{O}$ can effectively 
be taken to be the unit matrix,
$\mathbf{O} \cong {\bf 1}$.
This simplification 
is due to an additional $O(3)$ symmetry 
present in the lepton sector
when the neutrino mass 
spectrum is exactly degenerate \cite{BrancoQD}.

   Thus, up to corrections of the order of 
$\dma/(2m_\nu^2)$ and $\dms/(2m_\nu^2)$,
the matrix $\R$ in the expression for
$\ynu$, eq. (\ref{YnuQD}),
is effectively given by $e^{i\mathbf{A}}$.
The matrix $e^{i\mathbf{A}}$
can be explicitly 
calculated in terms of the three  
non-zero elements of $\mathbf{A}$. If we write 
\be
\mathbf{A}=\left(\ba{ccc} 0& a&b\\ -a&0&c\\ -b& -c&0 \ea\right)\,,
\ee
%
then
\be
e^{i\mathbf{A}}=
\mathbf{1}-\frac{\cosh r-1}{r^2}\mathbf{A}^2+i\frac{\sinh r}{r}\mathbf{A}\,,
\label{exp}
\ee
%
where $r=\sqrt{a^2+b^2+c^2}$. 

   Our final expression for $\ynu$
in the case of QD 
neutrino mass spectrum is
\be
\ynu \cong \frac{1}{v_u}M_R^{\sss 1/2}m_\nu^{\sss 1/2} 
e^{i\mathbf{A}}U^\dagger\,,
\label{yuk2}
\ee 
%
with $e^{i\mathbf{A}}$ given by (\ref{exp}). 

If we take the mixing angles 
$\theta_{12} = \theta_{\odot}$ and 
$\theta_{23} = \theta_{\rm A}$ as known
and neglect $\sin \theta_{13}$ in $U$, 
both 
$\ynu^\dagger\ynu$ and $\ynu\ynu^\dagger$
depend on 5 real parameters: 
$M_R$, $m_\nu$, $a$, $b$, $c$;
$\ynu^\dagger\ynu$ depends in addition on
the phases $\alpha$ and $\beta$.

   Yukawa couplings are expected to have 
moduli less than one, $|\ynu|\le 1$. 
Taking $a=b=c\equiv k$ we get from the 
diagonal elements of $\ynu$ the condition
\be
\cosh 3k\le \left|\frac{261\mbox{GeV}}{\sqrt{M_Rm_v}}-\frac 12\right|\,,
\ee
%
so that $k < \{1.4,0.9,0.3\}$ for $m_\nu=0.2\,$eV  
and $M_R=\{10^{10},10^{12},10^{14}\}$ GeV, respectively.

\vspace{-0.4cm}
\section{The processes $\ell_i\ra \ell_j + \gamma$} 
\label{dos}
\vspace{-0.2cm}

\hskip 1truecm The existence of two Yukawa 
couplings in the lepton sector generally 
causes lepton flavour violation in a way 
analogous \cite{SP76} to its quark sector counterpart 
in the Standard Theory. 
In the minimally extended Standard
Theory with massive neutrinos and
in the non-supersymmetric 
versions of the see-saw model, 
the decay rates and cross sections of the 
LFV processes are extremely
suppressed: one has, for example,
for the branching ratio 
of the $\mu\ra e+\gamma$ decay,
$BR(\mu\ra e+\gamma) < 10^{-47}$ \cite{SP76,lfv}. 
Such small branching ratios are unobservable. 
The present experimental limit is \cite{mega}
\be
BR(\mu\ra e+\gamma) < 1.2\times 10^{-11} .
\ee
%
This bound is expected 
to be improved at least by a few orders 
of magnitude in the future. 
In an experiment under preparation at 
PSI \cite{psi}, for instance, 
it is planned to reach a sensitivity to 
\be
BR(\mu\ra e+\gamma)\sim 10^{-14}\,.
\ee
%

 As we have seen, the rates of the LVF processes 
in the minimally extended Standard
Theory with massive neutrinos are so strongly
suppressed that these processes are 
not observable in practice.
In a SUSY theory the situation is very 
different because there is a new source 
of lepton flavor violation:
the soft SUSY breaking Lagrangian,
$\la_{soft}$. 
The breaking of SUSY will, 
generally, cause lepton flavor violation. 
Indeed, off-diagonal elements in the neutrino 
Yukawa coupling can give
rise to off-diagonal elements in the slepton 
mass matrix at low energies 
through renormalization group effects. 

  The slepton sector of the soft 
supersymmetry breaking Lagrangian has the form
\bea
-\la_{soft}&=&(\mathbf m_{\sss \s L}^2)_{ij} 
\s L_i^\dag\s L_j+(\mathbf m_{\s e}^2)_{ij}
\s e_{Ri}^*\s e_{Rj}+(\mathbf m_{\s \nu}^2)_{ij}
\s \nu_{Ri}^*\s \nu_{Rj}\nonumber \\
&&+\left(\mathbf{A}_{ij}^eH_d\s e_{Ri}^*\s L_j + 
\mathbf{A}_{ij}^\nu H_u \s\nu_{Ri}^*\s L_j+ h.c.\right)\,.
\eea
%
Lepton flavor violation
can be generated by
off-diagonal elements of the soft 
SUSY breaking parameters.  
The most conservative starting point for 
$\la_{soft}$ is the assumption of 
universality at the GUT scale M$_X$:
\bea
(\mathbf{m}^2_{\s L})_{ij}\hspace{-2mm}&=
&\hspace{-2mm}(\mathbf{m}^2_{\s e})_{ij}=
(\mathbf{m}^2_{\s\nu})_{ij}=\delta_{ij}m_0^2\,,\nonumber \\
\s m_{H_d}^2\hspace{-2mm}&=&\hspace{-2mm}\s m_{H_u}^2=m_0^2\,,\\
\mathbf{A}^\nu\hspace{-2mm}&=&\hspace{-2mm}\ynu a_0m_0,\quad \mathbf{A}^e=
\mathbf{Y}_ea_0m_0\, .\nonumber
\eea
%
Thus, at the unification scale, flavor is 
exactly conserved by $\la_{soft}$. Nevertheless, soft 
SUSY breaking terms suffer from renormalization 
via Yukawa and gauge interactions. In this way, 
LFV in the Yukawa couplings will induce LFV in 
the slepton mass matrices at low energy even if 
the slepton masses are flavour-universal at high energy.

  The RGE for the left-handed slepton mass matrix 
is given by (see, e.g., \cite{iba,JohnE})
\bea
\mu\frac{d}{d\mu}(\mathbf{m}_{\s L}^2)_{ij}&=
&\mu\frac{d}{d\mu}(\mathbf{m}_{\s L}^2)_{ij}
\Big|_{\sss \mathrm{MSSM}}+\frac{1}{16\pi^2}
\left[(\mathbf{m}_{\s L}^2\ynu^\dagger\ynu+
\ynu^\dagger\ynu\mathbf{m}_{\s L}^2)_{ij}\right.\nonumber\\
&&\left.+2(\ynu^\dagger\mathbf{m}_{\s\nu}^2\ynu+
\s m_{H_u}^2\ynu^\dagger\ynu +
\mathbf{A}_\nu^\dagger\mathbf{A}_\nu)_{ij}\right]\nonumber,
\eea
%
where the first term is the standard $\textrm{MSSM}$ 
term which has no LFV, while the second one is 
the source of LFV. In the leading-log approximation 
and with universal boundary conditions, 
the off-diagonal elements of the left-handed 
slepton mass matrix at low energy are given by
\beq
(\mathbf{m}_{\s L}^2)_{ij}\approx -\frac{1}{16\pi^2}(6+2a_0^2)m
_0^2(\ynu^\dagger\ynu)_{ij}\log \frac{M_X}{M_R}.\label{sle}
\eeq
%
This equation shows the connection 
between LFV in neutrino Yukawa couplings and 
LFV in slepton mass terms.

   Let us turn now to the 
lepton-flavor violating processes of the 
type $\ell_i\rightarrow \ell_j+\gamma$. 
The amplitude for this process  has the general form
\beq
T=\epsilon^\alpha\bar\ell_jm_{\ell_i}i
\sigma_{\alpha\beta}q^\beta(A_LP_L+A_RP_R)\ell_i \,,
\eeq
%
where $q$ is the momentum of the photon, 
$P_{R(L)}=(1 + (-) \gamma_5)/2$ and $A_L$ ($A_R$) is 
the coefficient of the amplitude when the 
decaying lepton is left-handed (right-handed). 
The corresponding branching ratio is 
\beq
BR(\ell_i\ra\ell_j+\gamma)=\frac{12\pi^2}{G_F^2}(|A_L|^2+|A_R|^2).
\eeq
%
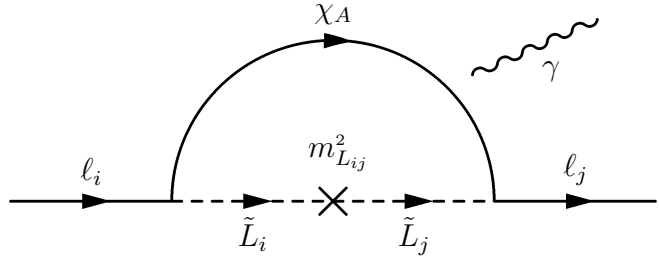
\begin{figure}
\bc
\begin{fmffile}{unos}
\begin{fmfgraph*}(250,100)
\fmfright{b,j,m,p}\fmfleft{c,i,k,q}\fmftop{t2}
\fmf{fermion,label=$\ell_i$}{i,o1}
\fmf{scalar,label=$\tilde{L}_{i}$,label.side=right}{o1,o3}
\fmf{scalar,label=$\tilde{L}_{j}$,label.side=right}{o3,o2}
\fmf{fermion,label=$\ell_j$,label.side=left}{o2,j}\fmffreeze
\fmf{phantom}{k,o4,o5,j}
\fmf{phantom}{q,o4,o5,p}\fmf{photon,label=$\gamma$}{o5,p}
\fmf{fermion,left,label=$\tilde{\chi}_A$,label.side=left}{o1,o2}
\fmfv{decoration.shape=cross,label=$m_{L_{\sss ij}}^{\sss 2}$,
l.angle=80,l.dist=0.4cm}{o3}
\end{fmfgraph*}
\end{fmffile}\ec
\vspace{-0.7cm}
\caption[no]{\footnotesize Feynman diagrams 
giving the dominant contribution 
to the $\ell_i\ra \ell_j+\gamma$ 
decay amplitude in the mass-insertion approximation. 
$\tilde\chi_A$ denotes charginos and neutralinos, and 
$\tilde L_i$ are the slepton doublets in the 
basis in which the gauge interactions and the 
charged-lepton Yukawa couplings are flavour-diagonal. 
The photon can be emitted 
from the chargino or the slepton lines.}
\label{feyn}
\end{figure} 
%
The terms $|A_{L,R}|$ contain the contributions 
of the neutralino and the chargino loops (see Fig. 1). 
Explicit expressions for $A_L$ and $A_R$ 
can be found in the literature \cite{his}. 
In the mass insertion approximation,
the diagrams contributing to 
$\ell_{i}\ra\ell_{j}+\gamma$ 
have the generic form shown in Fig. \ref{feyn} 
and the branching can be estimated using the
expression 
\beq
BR(\ell_i\ra\ell_j+\gamma)\simeq
\frac{12\pi^2}{G_F^2}|A_R|^2\simeq 
\frac{\alpha^3}{G_F^2}
\frac{|(\mathbf{m}^2_{\s L})_{ij}|^2}{m_S^8}\tan^2 \beta\,,
\eeq
%
where $m_S$ represents a scalar lepton mass. 
In the leading-log approximation, 
using (\ref{sle}), one finds \cite{iba,JohnE}
\beq
BR(\ell_i\ra\ell_j+\gamma)\simeq 
\frac{\alpha^3}{m_S^8G_F^2}\left|\frac{3+a_0^2}{8\pi^2}m
_0^2\log \frac{M_X}{M_R}\right|^2
\left|(\ynu^\dagger\ynu)_{ij}\right|^2\tan^2 \beta\,.
\eeq
%
Therefore, the off-diagonal elements of 
$\ynu^\dagger\ynu$ are the crucial 
quantities needed to estimate the branching ratios.

  Using the expression for $\ynu$ in
eq. (\ref{yuk2}) we find that in the case of 
QD neutrino mass spectrum and 
in the approximation of negligible splitting
between the neutrino masses,
\beq
(\ynu^\dagger\ynu)_{ij}\simeq
\frac{1}{v_u^2}M_Rm_\nu(Ue^{i2\mathbf{A}}U^\dagger)_{ij}\,.
\label{yc}
\eeq
%
For small values of $a$, $b$, and $c$, and negligible 
$s_{13}$ we obtain
\bea
(Ue^{i2\mathbf{A}}U^\dagger)_{21}\hspace{-3mm}&\simeq&\hspace{-3mm}
2i\left[-a(c_{12}^2 e^{i\am}+s_{12}^2e^{-i\am})c_{23} - 
e^{i\an }s_{23}(b c_{12}
+ c s_{12}e^{-i\am })\right] \label{21CR}\\
(Ue^{i2\mathbf{A}}U^\dagger)_{31}\hspace{-3mm}&\simeq&\hspace{-3mm}
2i\left[a(c_{12}^2 e^{i\am}+s_{12}^2e^{-i\am})s_{23} -
e^{i\an} c_{23}(b c_{12} + cs_{12}e^{-i\am})\right]\\
(Ue^{i2\mathbf{A}}U^\dagger)_{32}\hspace{-3mm}&\simeq&\hspace{-3mm}
2i\left[-2ia s_{12}c_{12}c_{23}s_{23}\sin \am + 
(bs_{12} - cc_{12}e^{i\am})(s_{23}^2e^{i\an}+c_{23}^2e^{-i\an})\right]\,
\label{32CR} 
\eea 
%
These elements 
control the $\mu\ra e+\gamma$, $\tau\ra e +\gamma$ and
$\tau\ra \mu+\gamma$ decay rates, respectively.

   Results for the LFV decay rates of interest
in the case of real $\mathbf{R}$ 
have been obtained, e.g., in \cite{iba,tan,DepPas}. 
We will analyze next the 
differences in the predictions for the
$\mu\ra e+\gamma$, $\tau\ra e +\gamma$,
$\tau\ra \mu+\gamma$ decay rates
which appear when the matrix $\mathbf{R}$ 
is complex. We will denote the matrix 
$\ynu^\dagger\ynu$ obtained by 
taking $\mathbf{R}$ real as $\ynu^\dagger\ynu|_\mathbb{R}$. 
One has \cite{iba,tan}
\bea
\left(\ynu^\dagger\ynu|_\mathbb{R}\right)_{ll'}\hspace{-3mm}&=
&\hspace{-3mm}\frac{M_R}{v_u^2}
\left[U_{l2}U_{l'2}^*(m_2-m_1)+U_{l 3}U_{l'3}^*(m_3-m_1)\right]\\
 &=&\hspace{-3mm}\frac{M_R}{v_u^2}
\left[U_{l2}U_{l'2}^*\frac{\dms}{2m_\nu}+
 U_{l3}U_{l'3}^*\frac{\dma}{2m_\nu}\right]\,,~l=\mu, l'=e~
{\rm and}~l=\tau,l'=e,\mu .
\label{21RR}
\label{ghj}
\eea 
%
This expression does not depend on 
$\R$ and has the special property that it 
depends on the masses only through the 
differences $m_2-m_1$ and $m_3-m_1$ 
but not through their absolute values. 
Thus, in contrast to eq. (\ref{yc}), 
there is no contribution 
proportional to $m_\nu$.
The main contributions 
in the cases of $\mu\ra e+\gamma$ and 
$\tau\ra e +\gamma$ decays
are of order $s_{13}\dma/(2m_\nu)$, 
or of order $\dms/(2m_\nu)$
if $s_{13}$ is rather small; in the case
of $\tau\ra \mu +\gamma$ decay
it is of order $\dma/(4m_\nu)$.
Therefore, as long as $|a|,|b|,|c| \gtap 10^{-3}$,
the  $\mu\ra e+\gamma$,
$\tau\ra e +\gamma$ and
$\tau\ra \mu +\gamma$ decay 
rates calculated 
using eq. (\ref{yc}), will typically 
be enhanced with respect to 
the rates calculated using eq. (\ref{ghj}), 
because in the case of 
QD neutrino mass spectrum one has
$m_\nu\gg \dma/(4m_\nu),
s_{13}\dma/(2m_\nu),\dms/(2m_\nu)$. 
The precise magnitude 
of this enhancement
depends on the values of the 
parameters $a$, $b$ and $c$,
contained in $\mathbf{A}$. 
  
  Expressions (\ref{21CR}) - (\ref{32CR}) 
and (\ref{21RR}) differ significantly
in one more aspect: in contrast 
to $\ynu^\dagger\ynu|_\mathbb{R}$,
eq. (\ref{21RR}), the quantity 
$\ynu^\dagger\ynu$ calculated
for complex $\mathbf{R}$ 
depends on the Majorana CP-violating phases
$\alpha$ and $\beta$ in the PMNS matrix
\footnote{This is valid also in the cases
of neutrino mass spectra with normal 
and inverted hierarchy \cite{JohnE}.}
$U$. Thus, the observation of the LFV processes
$\mu\ra e+\gamma$, $\tau\ra \mu +\gamma$, etc.
could allow one to get information about
these phases. Let us recall that
determining or even constraining 
the Majorana CP-violating phases in 
the neutrino matrix $U$ is a formidable problem 
(see, e.g., \cite{BPP1,PPW}). 

  In the numerical estimates which follow we take 
$s_{12} \equiv \sin\theta_{\odot}=0.6$, 
$s_{23} \equiv \sin\theta_{\rm A}=1/\sqrt{2}$, $\delta=0$, 
$\dma=3\times 10^{-3}$ eV$^2$, 
$\dms=7\times 10^{-5}$ eV$^2$, 
$m_\nu=0.3$ eV and consider two 
different values for $s_{13}$: $0;~0.2$. 
We have
\beq
\left|\left(\ynu^\dagger\ynu|_\mathbb{R}\right)_{31}\right|^2\sim
\left|\left(\ynu^\dagger\ynu|_\mathbb{R}\right)_{21}\right|^2
\simeq \frac{M_R^2m_\nu^2}{v_u^4}\times\left\{\ba{ll}6 \times 10^{-6} 
& \textrm{if } s_{13}=0.2, \\ 1.4 \times 10^{-8}&\textrm{if } s_{13}=0.0.
\ea\right.
\label{real}
\eeq
%
We will compare these results with 
the results we get for a complex 
matrix $\mathbf{R}$. 
We set $\alpha = \pi/2$ and $\beta = 0$.  
If we choose $|a|,|b|,|c| \simeq \mathcal{O}(10^{-1})$, 
we always get a result that is much larger 
than (\ref{real}). We have, for instance,
\beq
\left|\left(\ynu^\dagger\ynu\right)_{21}\right|^2
\simeq\frac{M_R^2m^2_\nu}{v_u^4}\times 
\left\{\ba{ll}0.34, & \textrm{for } (a,b,c)=(0.2,-0.4,0.5),\\ 
0.81,&\textrm{for }(a,b,c)=(0.4,0.3,0.2)\ea\right.,
\eeq
%
where we used eq. (\ref{yc}) with $s_{13}=0$
(the results for $s_{13}=0.2$ 
are only slightly different). We see that the 
coefficient in the right-hand side
of the above equation is 
always $\mathcal{O}(0.1-1.0)$. This result means 
that the branching ratio of the $\mu\ra e+\gamma$ 
decay will be enhanced with respect to the 
prediction based on eq. (\ref{real})~ 
\footnote{That if $\mathbf{R}$ is complex, 
$BR(\mu\ra e+\gamma)$ could be enhanced 
with respect to the branching 
ratio predicted for real $\mathbf{R}$, 
was noticed in \cite{iba}.}
approximately by a factor of 
$10^{5}$ to $ 10^{8}$  
depending on the value of $s_{13}$. 
Even if we take $|a|,|b|,|c|\simeq \mathcal{O}(10^{-2})$, 
there is still an enhancement of about
three to six orders of magnitude.
The same results are valid 
for the $\tau\ra e+\gamma$ decay rate.

 The $\tau\ra \mu+\gamma$ 
decay rate is also enhanced, 
but the magnitude of the enhancement 
is smaller than in the case 
of the $\mu\ra e+\gamma$ decay rate: by a factor of
$\sim 10^{4}$ for $|a|,|b|,|c|\simeq \mathcal{O}(10^{-1})$,
and of $\sim 10^{2}$ for 
$|a|,|b|,|c|\simeq\mathcal{O}(10^{-2})$.
This is due to the fact that the leading term in 
$\left(\ynu^\dagger\ynu|_\mathbb{R}\right)_{32}$
is not suppressed by $s_{13}$.

  Detailed predictions for
the rate of the $\mu\ra e +\gamma$ 
decay, obtained for real $\mathbf{R}$
and for QD neutrino mass spectrum, 
can be found in \cite{iba,tan,DepPas}.
They can be used,
together with eqs. (\ref{21CR}) and (\ref{21RR}), 
to estimate $BR(\mu\ra e +\gamma)$
for complex $\mathbf{R}$ we have considered.
The substantial enhancement 
we have found certainly makes
the importance of the 
searches for this decay
even more significant.

  Within the see-saw model,
the neutrino Yukawa couplings, $\ynu$,
plays a major role
in the generation of neutrino masses 
and in determining the rates of the 
LFV processes such as $\mu\ra e +\gamma$, 
$\tau\ra \mu+\gamma$ and $\tau\ra e +\gamma$ decays. 
It plays a fundamental role also in leptogenesis. 

\vspace{-0.4cm}
\section{The Leptogenesis Constraints}
\label{tres}
\vspace{-0.2cm}

\hskip 1.0cm The convenient dimensionless 
number which characterizes the magnitude of 
the baryon asymmetry of the Universe is the 
ratio of the baryonic charge density, 
$n_B-n_{\bar B}$, to the entropy density, $s$. 
The  presently observed baryon asymmetry is
\be
Y_B=\frac{n_B-n_{\bar B}}{s}=(0.1-1)\times 10^{-10}\,.
\ee
%
The aim of baryogenesis is to 
explain this number in terms of 
processes and fundamental parameters 
of particle physics. In leptogenesis, 
the out of equilibrium decays of heavy RH 
neutrinos produce a lepton asymmetry 
which is reprocessed by sphaleron 
processes into a baryon asymmetry.
%
%
If the light neutrinos are quasi-degenerate, 
\mbox{$m_\nu\gtap 0.1$} eV, 
the out-of-equilibrium condition 
cannot be satisfied and the 
amount of produced lepton asymmetry 
is strongly suppressed (see, e.g., \cite{plumZ,buch}),
unless the RH neutrinos are produced 
non-thermally. We shall consider 
leptogenesis via decays of RH neutrinos $N_{i}$ 
which are produced through inflaton decays \cite{kume}.

  At tree level the decay width of a 
heavy neutrino $N_i$ is,
\be
\Gamma_{Di}=\Gamma(N_i\ra H_u+l)+
\Gamma(N_i\ra H_u^c+l^c)=\frac{1}{8\pi}(\ynu\ynu^\dagger)_{ii}M_i\,.
\ee 
%
If \emph{CP} is not conserved by the neutrino Yukawa couplings, 
the interference between the tree and the one-loop diagram 
contributions to the $N_i$ decay amplitudes 
results in a lepton number production. 
The lepton number asymmetry per decay of a RH neutrino is
\bea
\epsilon_i&\equiv&\frac{\Gamma(N_i\ra H_u+l)-
\Gamma(N_i\ra H_u^c+l^c)}{\Gamma(N_i\ra H_u+l)+
\Gamma(N_i\ra H_u^c+l^c)}\nonumber \\ 
&\simeq&-\frac{1}{8\pi}\frac{1}{(\ynu\ynu^\dagger)_{ii}}
\sum_{j\neq i}\mathrm{Im}\left[\{(\ynu\ynu^\dagger)_{ij}\}^2\right]
\left[f(M_j^2/M_i^2)+g(M_j^2/M_i^2)\right]\,.\nonumber\\
\label{asy}\eea
%
Here $H_u$, $l$, and $N_i$ denote scalar or 
fermionic components of the 
corresponding supermultiplets, 
$f$ is the contribution from 
the one-loop vertex correction\cite{covi,Dort95,plumZ,buch}
\be
f(x)=\sqrt x\left[\log\left(\frac{1+x}{x}\right)\right]\,,
\ee
%
and $g$ is the contribution from the 
one-loop self energy diagrams, which can be 
reliably calculated in perturbation theory 
if the condition
\be
|M_i-M_j|\gg|\Gamma_i-\Gamma_j|
\label{cond}\ee
%
holds. One finds
\be
g(x)=\frac{2\sqrt x}{x-1}\,.
\ee
For quasi-degenerate RH neutrinos, $x\simeq 1$ and $g\gg f$.

 The ratio of the lepton number 
density $n_L$ to the entropy density $s$ 
produced by the inflaton decay is given by \cite{kume}
\be
\frac{n_L}{s}=\frac32\sum_i\epsilon_iBR(\phi\ra N_{i}N_{i})\frac{T_R}{m_\phi},
\ee
where $\phi$ denotes the inflaton field, 
$BR(\phi\ra N_{i}N_{i}) \equiv Br^{(i)}$ is the
$\phi\ra N_{i}N_{i}$ decay branching ratio,
$m_\phi$ is the mass of the inflaton
and  $T_R$ is the reheating temperature 
after the inflation. We have assumed that 
$M_R\ge T_R$ in order to prevent 
lepton-number violating processes 
from washing out the lepton asymmetry 
after the $N$'s have decayed. Part of the 
lepton asymmetry is immediately converted 
into baryon asymmetry via the sphaleron effect,
\be
\frac{n_B}{s}=C\frac{n_L}{s},
\ee 
%
with $C\simeq -0.35$ in the MSSM.

 Using eq. (\ref{yuk}) we obtain
\be
\ynu\ynu^\dagger=\frac{1}{v_u^2}D_M^{\sss 1/2}
\mathbf{R}D_m\mathbf{R}^\dagger D_M^{\sss 1/2}\,.\label{hg}
\ee 
%
Hence, in general, leptogenesis is 
independent of the mixing angles and 
phases contained in the PMNS matrix $U$. 
If $\R$ is real, $\mbox{Im}(\ynu\ynu^\dagger)=0$ 
and leptogenesis cannot work. This is a model 
independent statement and it is the main 
reason we have to assume that $\R$ is complex.

  For quasi-degenerate neutrinos, 
eq. (\ref{hg}) can be further simplified,
\be
\ynu\ynu^\dagger \simeq \frac{M_Rm_\nu}{v_u^2}e^{i2\mathbf{A}}\,.
\ee
%
It is well-known that if the RH neutrinos 
are completely degenerate, the generated 
lepton asymmetry is zero \cite{pil}. Thus, one 
has to break the exact degeneracy in the heavy 
RH neutrino masses. We write
\be
M_2=M_1(1-\varepsilon_2),\quad M_3=
M_1(1-\varepsilon_3),\quad |\varepsilon_3|\gg |\varepsilon_2| \,.\label{deg}
\ee
%
The maximal values of $|\varepsilon_{2}|$ and $|\varepsilon_{3}|$ 
which are naturally consistent with a 
low-energy quasi-degenerate neutrino mass spectrum 
are $|\varepsilon_{2}|\simeq \Delta m_\odot^2/2m_\nu^2$ 
and $|\varepsilon_{3}| \simeq \dma/2m_\nu^2$.

Conditions (\ref{cond}) for small $r$ translate into
\bea
|\varepsilon_2|&\gg&\frac{1}{4\pi}|c^2-b^2|\,\frac{M_R}{10^{14}\mbox{GeV}}\,
\label{e2} ,\\
|\varepsilon_3|&\gg&\frac{1}{4\pi}|c^2-a^2|\,\frac{M_R}{10^{14}\mbox{GeV}}\,
\label{e31} ,\\
|\varepsilon_3|&\gg&\frac{1}{4\pi}|b^2-a^2|\,\frac{M_R}{10^{14}\mbox{GeV}}\,
\label{e32} ,
\label{gg}\eea
%
where we have used $m_\nu=0.3\,$eV and $v_u=174\,$GeV. 
Since $|\varepsilon_2|\simeq 10^{-4}$ and 
$|\varepsilon_3|\simeq 10^{-2}$, 
for the extreme value  $M_R\simeq 10^{14}$GeV 
these conditions are satisfied as long as 
$|b|,|c| \ltap 10^{-2}$ and $|a| \ltap 10^{-1}$. 
For $M_R\simeq 10^{10}$GeV, eqs. (\ref{e2}) - (\ref{e32})
lead to the constraints
$|a|,|b|,|c| \ltap 1$. 

   We shall compute next the lepton number 
asymmetries, eq. (\ref{asy}). From (\ref{exp}) we get
\bea
\mathrm{Im}\left[(e^{i2\mathbf{A}})_{\sss 12}
(e^{i2\mathbf{A}})_{\sss 12}\right]&=&
2\frac{abc}{r^3}\sinh 2r(\cosh 2r-1)\nonumber \\
&=&\mathrm{Im}\left[(e^{i2\mathbf{A}})_{\sss 23}
(e^{i2\mathbf{A}})_{\sss 23}\right]\nonumber\\
&=&\mathrm{Im}\left[(e^{i2\mathbf{A}})_{\sss 31}
(e^{i2\mathbf{A}})_{\sss 31}\right]\label{hyp}
\eea
%
and $\mathrm{Im}\left[(e^{i2\mathbf{A}})_{\sss ij}(e^{i2\mathbf{A}})_{\sss ij}\right]=-\mathrm{Im}\left[(e^{i2\mathbf{A}})_{\sss ji}(e^{i2\mathbf{A}})_{\sss ji}\right]$.
Assuming that $a$, $b$ and $c$ are small, 
we can expand the hyperbolic functions in (\ref{hyp}) to obtain
\bea
\epsilon_1\hspace{-2mm}&=&\hspace{-2mm}\epsilon_2
\simeq \frac {1}{\pi}\frac{M_Rm_\nu}{v_u^2}\frac{abc}{\varepsilon_{2}}\,\\
\epsilon_3\hspace{-2mm}&\simeq&
\hspace{-2mm}-\frac{2}{\pi}\frac{M_Rm_\nu}{v_u^2}\frac{abc}{\varepsilon_{3}}\,
\eea
%

 The baryon asymmetry thus generated is 
\be
\frac{n_B}{s} \simeq 1.4\times 10^{-8}\left(\frac{2M_R}{m_\phi}\right)\!
\left(\frac{T_R}{10^{8}GeV}\right)\!\left(\frac{m_\nu}{0.1eV}\right)\!
\left(-\frac{abc}{\varepsilon_2}\right)\!(B_r^{(1)}+B_r^{(2)}),
\ee 
%
where we have neglected the $\epsilon_3$ 
contribution.
Hence, the empirical baryon asymmetry 
is obtained with a reheating temperature 
of $T_R\simeq 10^{8}$GeV for 
$|abc/\varepsilon_2|\simeq 10^{-1}$ and a natural 
choice of the remaining parameters. 
Since $|\varepsilon_2|\simeq 10^{-4}$, we have 
\be
|abc|\simeq 10^{-5}\,.
\label{lep}
\ee
%
Larger values of $|abc|$ are possible if $2M_R/m_\phi\ll 1$. 
Higher reheating temperatures would be compatible with 
smaller values of the product 
$|abc|$, but would also lead to the cosmological
gravitino problem \cite{gra}. 

The baryon asymmetry is proportional to 
the product $abc$ and therefore none of 
the three parameters 
can be zero. Moreover, they cannot 
be exceedingly small, otherwise 
it would be impossible to 
reproduce the baryon asymmetry. 
Equation (\ref{lep}) implies that at least one 
of them - $|a|$, $|b|$, or $|c|$, must be of 
order $10^{-2}$ or larger. This number 
fixes the possible enhancement of 
$BR(\mu\ra e +\gamma)$: 
about four orders of magnitude for $s_{13}=0.2$ 
and six orders of magnitude for $s_{13}=0$. 
$BR(\tau\ra e +\gamma)$ is 
enhanced by similar factors,
while $BR(\tau\ra \mu +\gamma)$
is enhanced approximately by two
orders of magnitude. 

\vspace{-0.2cm}
\section{Conclusions}
\vspace{-0.2cm}

\hskip 1.0cm   We have considered the 
$\mu\ra e + \gamma$,
$\tau\ra \mu + \gamma$ and
$\tau\ra e + \gamma$
decay branching ratios in a
class of SUSY GUT models 
with see-saw mechanism of
neutrino mass generation.
We have assumed
that the orthogonal $\mathbf{R}$ matrix 
which was introduced in \cite{iba}
and which is related to the neutrino Yukawa
coupling matrix $\ynu$, is complex.
This is required in order for 
the model to be compatible with the
leptogenesis scenario of 
generation of the baryon asymmetry.
In this case $\mathbf{R}$
can be represented as 
$\mathbf{R} = e^{i\mathbf{A}}\mathbf{O}$,
where $\mathbf{A}$ 
and $\mathbf{O}$ are respectively real 
antisymmetric and real orthogonal
matrices. The matrix $\mathbf{A}$ 
can be parametrized 
by 3 real parameters,
$a$, $b$ and $c$.
We have considered the
case of quasi-degenerate spectrum 
of light neutrinos, 
$m_{1,2,3} \cong m_{\nu}$, 
$m^2_{\nu} >> \dma,\dms$,
where $\dma$ and $\dms$
are the neutrino mass-squared differences
which drive the atmospheric and solar neutrino 
oscillations. 
Assuming that the 
heavy right-handed neutrinos
are also quasi-degenerate in mass,
$M_{1,2,3} \cong M_R$,
and that the soft SUSY breaking 
slepton mass terms are flavour-universal at 
the GUT scale, we have 
derived approximate expressions for  
$\mu\ra e + \gamma$, 
$\tau\ra \mu + \gamma$ and 
$\tau\ra e + \gamma$ decay rates.
Apart from the standard SUSY soft-breaking 
parameters
($m_0$, $a_0$, $\tan\beta$, $m_{1/2}$), 
the decay rates depend on $m_{\nu}$, $M_R$,
on the mixing angles $\theta_{12}$ and
$\theta_{23}$ which control the solar and 
atmospheric neutrino oscillations,
on the Majorana CP-violating phases
in the PMNS mixing matrix $U$
and on the parameters $a$, $b$ and $c$.
We have found that for complex $\mathbf{R}$, 
the branching ratios of the 
indicated LFV decays 
are considerably larger than 
when $\mathbf{R}$ is taken to be real:
for $a \sim b \sim c \sim 10^{-1}$, for instance,
$BR(\mu\ra e + \gamma)$ and
$BR(\tau\ra e + \gamma)$ are enhanced
approximately by a factor 
of $10^{5}$ to $10^{8}$ with respect
to the case of real $\mathbf{R}$,
while $BR(\tau\ra \mu + \gamma)$
is enhanced by approximately
four orders of magnitude.
We used the model of leptogenesis 
with light quasi-degenerate neutrinos,
in which the heavy RH Majorana neutrinos 
are assumed to be produced 
non-thermally in the inflaton decay,
to get  constraints on $a$, $b$ and $c$. 
The baryon asymmetry 
is proportional to the product
$abc$ of the three parameters
associated with the complexity
of $\mathbf{R}$. 
For values of the
RH neutrino mass
$M_R$ characteristic
for the leptogenesis model,
the observed asymmetry
can be reproduced 
for $|abc| \sim 10^{-5}$.
If $BR(\mu\ra e+\gamma)$ 
and $BR(\tau\ra e+\gamma)$
are evaluated for values of
$a$, $b$ and $c$ compatible
with the leptogenesis constraint, 
the enhancement we found 
is approximately by a factor of
$10^{3}$ and $10^{6}$
for values of 
$\sin^2\theta_{13} = 0~{\rm and}~ 0.04$.   
The corresponding 
enhancement of $BR(\tau\ra \mu +\gamma)$ 
is approximately by two orders of magnitude.

 Besides $a$, $b$ and $c$, 
and the Majorana CP-violating phases
in the PMNS matrix $U$,
$BR(\mu\ra e+\gamma)$
depends also on SUSY soft breaking 
parameters ($m_0$, $a_0$, 
$\tan\beta$, $m_{1/2}$) and the 
RH neutrino mass $M_R$. 
Given the existing experimental bound on 
$BR(\mu\ra e+\gamma)$,
our results can be used, in particular,
to further constrain 
the space of the supersymmetric 
parameters in the case of quasi-degenerate
neutrino mass spectrum. This requires 
a more detailed numerical analysis 
which is beyond
the scope of the present work.

  If neutrinos will be proven
experimentally to have
a quasi-degenerate mass spectrum
and the neutrino masses are generated
via the see-saw mechanism
within a SUSY GUT theory, 
the process $\mu\ra e+\gamma$ 
should be observable in the planned experiments 
of the next generation provided the
supersymmetric particles have masses
in the range of several hundred GeV.
Additional constraints from 
data on the $\tau\ra \mu+\gamma$ and 
$\tau\ra e+\gamma$ decays 
and from leptogenesis 
can be used to determine 
the matrix $\mathbf{A}$ and the
RH neutrino mass $M_R$. With that information, 
the neutrino Yukawa couplings $\ynu$ could 
be almost fully reconstructed. 

\vspace{0.2cm}
{\bf Acknowledgements.} We would like to thank
W. Rodejohann for useful discussions.
S.P. would like to thank SISSA for very kind hospitality
during the completion of the present work.
This work was supported in part by 
the EC network HPRN-CT-2000-00152, 
by the Italian MIUR under the program 
``Fenomenologia delle Interazioni Fondamentali'' (S.T.P.)
and by the U. S. Department of Energy (S.P.).

\end{document}